\documentclass{aa}  

\usepackage{graphicx}
\usepackage{txfonts}
\usepackage{lipsum}

\usepackage{lscape}             
\usepackage{placeins}           
\usepackage[table,xcdraw]{xcolor}
\usepackage{bm}
\usepackage{hyperref}
\hypersetup{
    colorlinks=true,
    linkcolor=blue,
    filecolor=magenta,      
    urlcolor=cyan,
    linkcolor=blue,  
    citecolor=blue,        
}

\usepackage{orcidlink}
\usepackage{tabularx}
\usepackage{multirow}
\usepackage[final]{microtype}

\let\orgautoref\autoref
\providecommand{\Autoref}[1]{\def\equationautorefname{Equation}
\def\figureautorefname{Figure}\def\sectionautorefname{Section}\def\subsectionautorefname{Section}\def\subsubsectionautorefname{Section}\def\tableautorefname{Table}\orgautoref{#1}}
\renewcommand{\autoref}[1]{\def\equationautorefname{Eq.}
\def\figureautorefname{Fig.}\def\sectionautorefname{Sect.}\def\subsectionautorefname{Sect.}\def\subsubsectionautorefname{Sect.}\def\tableautorefname{Table}\orgautoref{#1}}

\begin{document}

   \title{\texttt{argosim}: a Python package for radio interferometric simulations}

   \subtitle{}

   \author{Ezequiel Centofanti\inst{1}\fnmsep\thanks{E-mail: ezequiel.centofanti@cea.fr}\orcidlink{0009-0001-8461-8451},
        Emma Ayçoberry \inst{1}\orcidlink{0000-0002-9235-1195},
        Samuel Farrens \inst{1}\orcidlink{0000-0002-9594-9387},
        Samuel Gullin\inst{2}\orcidlink{0000-0003-2491-2883},
        Manal Bensahli\inst{1},
        Jean-Luc Starck\inst{1,2}\orcidlink{0000-0003-2177-7794},
        John Antoniadis\inst{2,3}\orcidlink{0000-0003-4453-3776}
        }

   \institute{
        Université Paris-Saclay, Université Paris Cité, CEA, CNRS, AIM, 91191, Gif-sur-Yvette, France
        \and
        Institutes of Computer Science and Astrophysics, Foundation for Research and Technology Hellas (FORTH), GR-70013 Heraklion, Greece 
        \and
        Max-Planck-Institut für Radioastronomie, Auf dem Hügel 69, DE-53121 Bonn, Germany}

   \date{Received August, 2025}
 
  \abstract
    {In this paper, we present \texttt{argosim}, a Python package for simulating radio interferometric observations. The \texttt{argosim} package is modular, lightweight and compatible with all major operating systems. Its computational backend is written in \texttt{JAX}, which allows for greatly accelerated performance as well as the advantage of being fully differentiable. We detail the main \texttt{argosim} modules and describe how to use them to generate an observation, from the antenna positions to the cleaned image. The package is a fully open-source project, and its code is publicly available on GitHub. \href{https://github.com/ARGOS-telescope/argosim}{\includegraphics[width = 0.017 \textwidth]{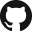}}}


\titlerunning{\texttt{argosim}}
\authorrunning{E. Centofanti et al.}
\maketitle
\nolinenumbers

\section{Introduction}
Modern radio interferometers deploy thousands of antennas that collect data at high temporal and spectral resolution, resulting in billions of $uv$-samples per observation \citep{dsa2000, ngVLA, HIRAX, ska2009}, and reaching data rates of several Tbits/s. 
This volume of data presents significant processing challenges, requiring substantial computational resources, and is nearly impossible to store in its raw form. For example, the Square Kilometre Array (SKA) \citep{ska2009} will be forced to discard the majority of its raw data, yet it will still need to archive over several hundred petabytes of processed data per year. 
Given the unprecedented volume and rate of collected data, it is essential to optimise image reconstruction algorithms to the point that observations can be processed in near real time.
Furthermore, since raw data will not be stored, all analysis steps must be as accurate and robust as possible, as reprocessing will not be feasible. 

\sloppy
As the next generation of radio interferometers moves through their design and construction phases, simulation tools are essential for developing and testing methods in image reconstruction, calibration, data processing, and storage. Several software packages exist for processing interferometric data \citep{CASA_2022,offringa-wsclean-2014,meqtrees2010,Miriad95,oskar,rascil}, with one of the most widely used being CASA (Common Astronomy Software Applications for Radio Astronomy), originally developed for the Atacama Large Millimeter/submillimeter Array (ALMA) \citep{alma} and Very Large Array (VLA) \citep{vla1983, vla1980} telescopes. CASA is written in C++ and features a Python-based interface. It is a comprehensive suite that offers tools for imaging, calibration, data reduction and image deconvolution. CASA is well supported by the radio astronomy community and receives regular updates.
\fussy

Although highly capable, CASA may be less suitable for lightweight or experimental workflows as it can be resource-intensive, requiring significant CPU and memory. Its installation can also be challenging due to the large number of dependencies and the fact that it is written in C++, especially on machines with newer CPU architectures.
Furthermore, its steep learning curve and reliance on multiple configuration files can pose barriers to quick prototyping and streamlined simulation tasks.
To address the need for lightweight and flexible simulation tools, we developed \texttt{argosim}, a Python package for radio interferometric simulations. This software was created in the context of the ARGOS project\footnote{\url{https://argos-telescope.eu/}}, however, it remains generalisable to many other radio interferometers and can be used for different science cases. 

ARGOS is a European initiative aimed at designing a cutting-edge, low-cost, and sustainable radio interferometer in Crete, Greece. The project envisions the deployment of an instrument composed of approximately one thousand antennas, operating in the 1-3 GHz frequency range with baselines extending up to 4 km. 
It will address a wide range of science cases, including the detection of fast radio bursts (FRBs), pulsars, and gravitational waves (GWs). 
ARGOS is expected to detect a few strongly lensed FRBs per year, opening new avenues for precision cosmology, such as enabling a 1\% measurement of the Hubble constant $H_0$ \citep{2021A&A...645A..44W}. 
Additionally, ARGOS will detect and characterise radio transients through imaging, providing insight into accreting compact objects, the nature of supernova progenitors and the physics of cosmic explosions.
The ARGOS pathfinder, currently under construction, consists of five six-meter parabolic antennas and serves as a proof-of-concept for both the hardware and software of ARGOS. 
An imaging example with the ARGOS pathfinder is presented in \autoref{sec:practical}.

In this paper, we present the first release of \texttt{argosim}, an open-source, lightweight and modular interferometric simulation package written in Python with a \texttt{JAX} computational backend. This paper is structured as follows. The following section introduces the fundamentals of radio interferometric imaging. \Autoref{sec:software} outlines the software design principles, with a focus on accessibility, reproducibility, and maintainability. \Autoref{sec:pipeline} presents the \texttt{argosim} simulation pipeline, from antenna configuration to the generation of synthetic observations, and the computation of image reconstruction metrics. \Autoref{sec:jax} discusses the advantages of accelerated array computation within \texttt{argosim}. \Autoref{sec:future} outlines plans for future development, and the paper concludes with final remarks in \autoref{sec:conclusions}.

\section{Radio interferometric imaging}
\label{sec:interferometry}
Radio interferometers correlate signals between pairs of antennas, which by the principle of interferometry, measure the magnitude of sinusoidal spatial components of the observed sky. These components are called visibilities and their relation to the observed sky is given by the van Cittert-Zernike theorem \citep{Thompson_interferometry}.

\begin{equation}
\label{eq:vCZ}
    \mathcal{V}(u,v,w) = \iint\frac{i_{{\rm sky}}(l,m)}{n} e^{-2\pi j(ul+vm+w(n-1))}{\rm d}l{\rm d}m,
\end{equation}
where $\mathcal{V}(u,v,w)$ is the complex visibility function, $i_{{\rm sky}}$ is the observed sky intensity as a function of the two direction cosines $(l, m)$ and $n$ is the third direction cosine
\begin{equation}
    n = \sqrt{1-l^2-m^2}.
\end{equation}
If we assume that the observed sources do not subtend large angles in the sky, the small sky approximation is fulfilled
\begin{equation}
    l^2+m^2 \rightarrow 0,
\end{equation}
which simplifies \autoref{eq:vCZ} yielding
\begin{equation}
\label{eq:vCZ_2d}
    \mathcal{V}(u,v) = \iint i_{{\rm sky}}(l,m) e^{-2\pi j(ul+vm)}{\rm d}l{\rm d}m.
\end{equation}
Neglecting the $w$ term introduces a distortion associated with projecting the celestial sphere onto a plane \citep{1999ASPC..180.....T}. Nevertheless, if the imaged region is sufficiently small and centred on the direction of observation, this effect can be ignored \citep{2008ISTSP...2..647C}.
We can rewrite \autoref{eq:vCZ_2d} in terms of the two-dimensional (continuous) Fourier transform
\begin{equation}
    \mathcal{V}(u,v) = \mathcal{F}\left[ i_{{\rm sky}}(l,m) \right](u,v).
\end{equation}
However, since each pair of antennas samples the complex visibility function at a specific position in the $uv$-plane, which depends on their separation, known as the baseline, only a limited number of noisy $uv$-points are measured 
\begin{equation}
   \mathcal{V}_{i} = I_{{\rm sky}}(u_i, v_i) \;+\; N(u_i,v_i) .
\end{equation}
If we represent the set of $uv$-sampling positions by a mask in Fourier space, we can describe the measured visibilities as follows:
\begin{equation}
\label{eq:vis}
    V = M \;\odot\; \left(F \,i_{{\rm sky}} + N\right),
\end{equation}
where $V$ are the measured visibilities, $F$ is the discrete Fourier transform, $M$ is the sampling mask (containing ones at $uv$-positions with samples and zeros elsewhere), $\odot$ is the Hadamard product which represents the element-wise matrix multiplication, and $N$ is the additive Gaussian noise. Note that the noise, which is white in $uv$-space, is correlated through the sampling mask in real space. 

\subsection{Aperture synthesis}
\label{sec:apsyn}
        \Autoref{eq:vis} defines the observational model in radio interferometry, in which a finite number of $uv$-samples, defined by the sampling mask $M$, are measured.
        Each $uv$-point corresponds to a measurement of a spatial component of the observed sky.
        The frequency and orientation of the measured components depend on the relative location of each pair of antennas. The vector of separation between a pair of antennas is called the baseline and has units of metres (usually measured in kilometres). A radio interferometer with $N$ antennas contains a total of $N(N-1)/2$ baselines.

        The spatial components measured by each pair of antennas are not determined by their physical separation alone, but rather by their relative positioning with respect to the direction of the observed source. Just as a circular disc appears elliptical when viewed in perspective, the apparent distance between antennas varies depending on the observation angle. For this reason, the baselines are converted to the Cartesian coordinate system $(u,v,w)$, where the $uv$-plane is perpendicular to the line of sight and the $w$-axis is aligned with it.
        The interference phenomenon is linked to the phase shift of the electromagnetic (EM) waves due to the optical path difference (OPD) between the source and each antenna. For a given pair of antennas and a direction of observation, the OPD is fixed in metres, but the phase delay (difference in cycles) depends on the wavelength of the incoming EM wave. Consequently, in the $uvw$-space, baselines are usually expressed relative to the operating wavelength of the radio telescope $\lambda$ (usually in units of k$\lambda$, that is, $10^3 \lambda$).
        
        Finally, a radio interferometric observation can involve observing an area of the sky over a period of time, which is known as tracking. Due to the rotation of the Earth with respect to the celestial sphere, the direction of the source, and thus the $uv$-points, are different for each time-step \citep{10.1093/mnras/125.1.39}. This allows more spatial frequencies ($uv$-points) to be sampled. Furthermore, radio telescopes do not operate on a single frequency, but on a frequency band that can be subdivided into several spectral bins, which increases the $uv$-coverage \citep{1994A&AS..108..585S} resulting in a total of $N(N-1)/2\times N_{\rm timesteps}\times N_{\rm freqs}$ $uv$-samples for an $N$-antenna interferometer.

    \subsection{Deconvolution}
        By applying the inverse Fourier transform to \autoref{eq:vis} we can obtain the observation, also called dirty image, which can be expressed as the convolution of the sky intensity and the synthesised beam of the radio interferometer $b$:
        \begin{equation}
            i_{\text{obs}} = b \;\ast\; i_{{\rm sky}} \;+\; n,
            \label{eq:obs_model_real}
        \end{equation}
        where $b = \mathcal{F}^{-1}M$, $\ast$ is the convolution operator and $n$ is coloured noise. We note that, analogous to an optical system, the observed image is the result of the convolution of a 2D function, $b$, with the real sky image $i_{sky}$. The function $b$ represents the point spread function (PSF) of the telescope, which in the context of radio interferometry is known as the dirty beam.
        
        Recovering the underlying sky intensity from the measured visibilities (or the dirty sky) is an inverse problem, which has no closed solution. To solve this type of problem, iterative algorithms are often used in conjunction with certain assumptions on the properties of the observed sources. One of the most widely used deconvolution algorithms is CLEAN \citep{1974A&AS...15..417H}, and its subsequent variants such as Clark's Clean method \citep{1980A&A....89..377C}, Cotton-Schwab Clean \citep{1984AJ.....89.1076S}, and Multi-Scale Clean \citep{4703304}. There exist other more recent methods that use compressed sensing and sparse representation techniques such as SASIR \citep{2015JInst..10C8013G, garsden2015}, and methods that use neural networks and generative diffusion models \citep{dia2025irisbayesianapproachimage}.

\hfill

The \texttt{argosim} package enables simulation of the measurement process from the sampling mask, which depends on the position of the antennas, to the generation of the sky model, sampling of the visibilities, and image deconvolution using the simple CLEAN algorithm. In \autoref{sec:pipeline} we give more details about the \texttt{argosim} pipeline and how to simulate a radio interferometric observation.

\section{Software design}
\label{sec:software}
    
    \subsection{Open-source development}

    The \texttt{argosim} Python package is fully open-source, and the code is publicly available on the GitHub organisation of the ARGOS consortium\footnote{\url{https://github.com/ARGOS-telescope/argosim}}. The repository also contains evolving documentation, which includes a detailed description of every \texttt{argosim} module and function, as well as tutorials and usage examples (see also \autoref{sec:practical}).
    We actively encourage contributions from the broader community via pull requests and issue submissions. To streamline communication, we provide issue templates that support structured reporting of bugs, feature requests, assistance inquiries, and other encountered problems. 
    
    \subsection{Continuous integration}

    To ensure both short- and long-term reliability, we have implemented continuous integration using GitHub Actions. Each pull request triggers a suite of tests on macOS and Ubuntu environments to validate new features and verify the stability of existing functionality.
    The continuous integration framework includes unit tests for the core modules, implemented with the \texttt{pytest} \citep{pytest} Python package. Current test coverage reaches $94\%$, providing a high level of code robustness. Additionally, the pipeline incorporates style checks to enforce code consistency—using the \texttt{Black} \citep{black} and \texttt{isort} \citep{isort} packages—as well as documentation builds to guarantee that the online API documentation remains up to date.
    This setup facilitates early detection of bugs and provides immediate feedback to developers. It also ensures that all changes conform to the established standards before being merged into the main codebase.

   \subsection{Reproducibility}

    As part of the \texttt{argosim} package release, we have implemented several measures to ensure the reproducibility and robustness of results. First, the accompanying documentation includes a detailed list of required Python packages, along with the specific versions when relevant. This allows users to obtain consistent results across different platforms and environments.
    Second, we provide a pre-built Docker image\footnote{\url{https://ghcr.io/argos-telescope/argosim:main}} that users can pull and run. This container enables users to execute the code interactively without manually installing dependencies, thereby avoiding potential compatibility or installation issues—only Docker\footnote{\url{https://www.docker.com/}} is required.
    These measures are designed to promote both reproducibility and usability, ensuring that users can reliably reproduce current and future results while simplifying installation and execution.

\section{Radio interferometry pipeline} 
\label{sec:pipeline}
    The \texttt{argosim} simulation package is modular, and each of the modules are related to different stages in the generation of radio interferometric observations. In this section, we detail the main modules available in \texttt{argosim}, their content and give some practical examples. Further details are available in our extensive documentation\footnote{\url{https://argos-telescope.github.io/argosim/}}. A list of the third-party software packages used in \texttt{argosim} is presented in \autoref{app:dependencies}, and a list of the main modules is provided in \autoref{app:modules}.
    
    \subsection{Interferometric array}

        \begin{figure}
            \centering
            \includegraphics[width=\linewidth]{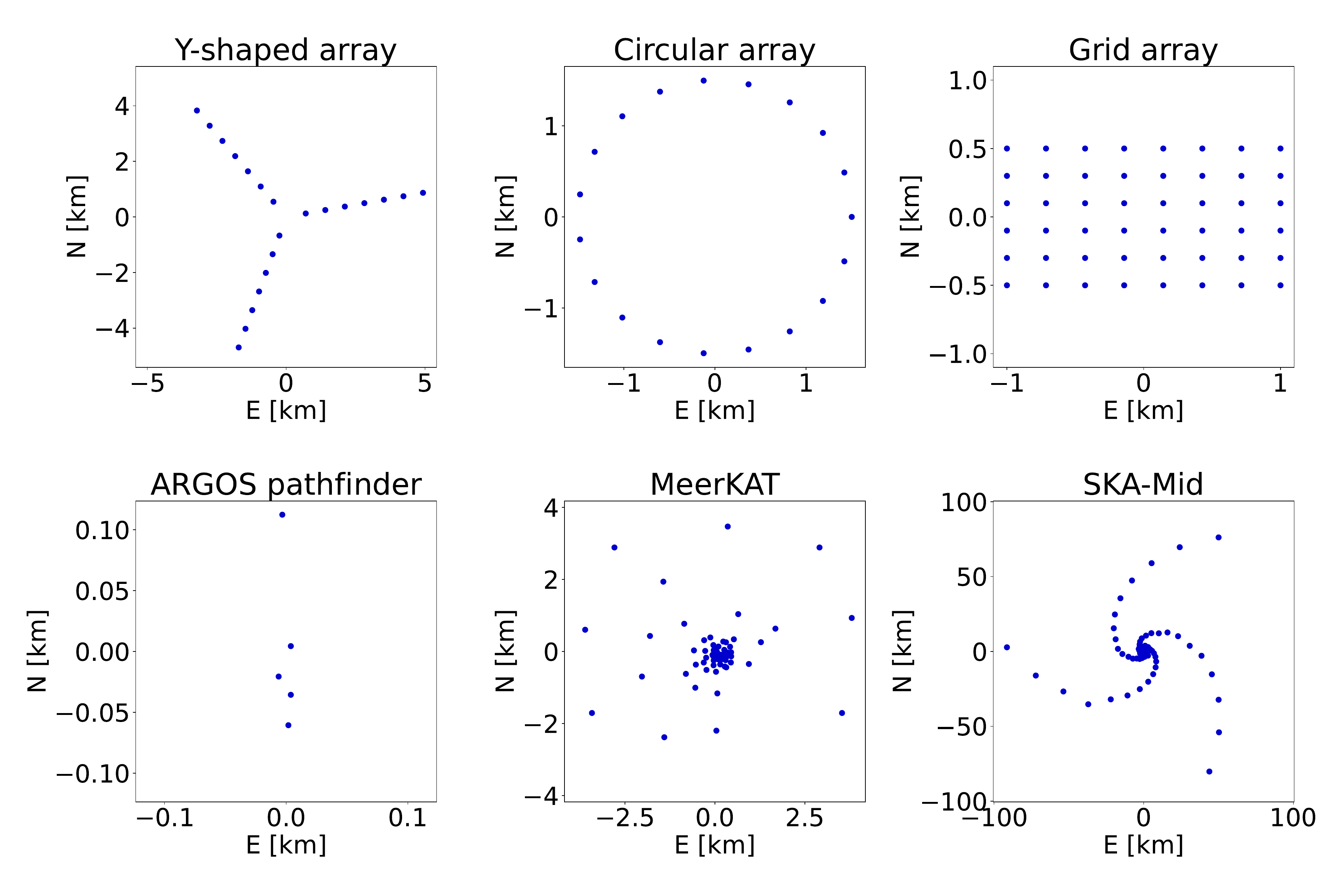}
            \caption{Simulated array configurations in the East North plane. Top: example of \texttt{argosim} parametrisable arrays, including Y-shaped, circular and regular-grid antenna configurations. Bottom: user-defined arrays, featuring the ARGOS pathfinder, the MeerKAT radio telescope and the SKA-Mid antenna layout.}
            \label{fig:arrays}
        \end{figure}
        
        The \texttt{antenna\_utils} module provides functionality related to the configuration and properties of the antenna array. We represent a radio interferometer as a list of antenna locations. The position of each antenna is given in East North Up (ENU) coordinates with respect to the centre of the array, defined by the latitude and longitude of the telescope on Earth.
    
        This module has several parametrisable functions that allow the generation of arrays with the following geometric distributions: Y-shaped array, circular array, 2D uniform grid array, and 3D random antenna placement. In the latter case, the third (Up) coordinate corresponds to the elevation above sea level of each antenna, which allows non-coplanar arrays to be simulated. Any of the generated arrays can be saved in a configuration file that can be loaded later, ensuring reproducibility of the simulations. It is also possible to define and load custom arrays from text files listing the ENU or (latitude, longitude, elevation) coordinate positions of each antenna.

        \Autoref{fig:arrays} shows some array configurations simulated with \texttt{argosim}. The first row shows the parametrisable Y-shaped, circular, and regular-grid antenna configurations. The second row shows three custom arrays: the planned antenna distribution for the ARGOS pathfinder, the MeerKAT radio telescope configuration, and the SKA-Mid antenna layout \citep{SKA-TEL-SKO}.
        
    \subsection{\texorpdfstring{$uv$}{TEXT}-samples}
\begin{figure}
            \centering
            \vspace{.15cm}
            \includegraphics[width=\linewidth]{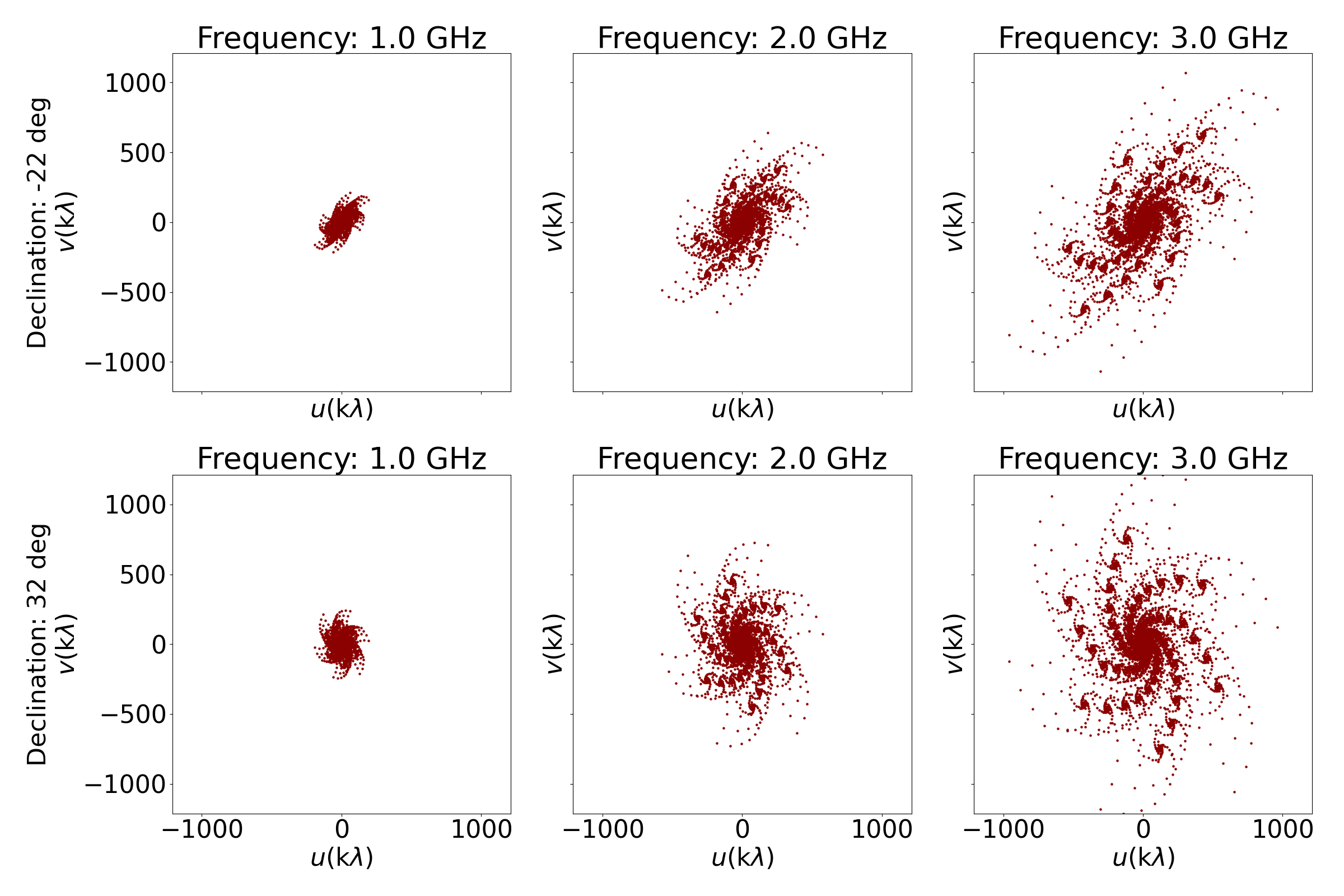}
            \caption{Simulated snapshot $uv$-coverage of the SKA-Mid array layout for three different working frequencies and two different observation declinations.}
            \label{fig:baselines}
        \end{figure}

        As mentioned in \autoref{sec:apsyn}, the aperture synthesis process allows us to obtain the $uv$-sampling points given the antenna positions, the latitude of the array, the source declination, the operating frequency of the antennas, and the observation time. The \texttt{argosim} package allows simulation of $uv$-coverage under all these circumstances.
        \Autoref{fig:baselines} shows simulations of the SKA-Mid array snapshot (without time integration) $uv$-coverage for three different operating frequencies: 1, 2, and 3 GHz. We remark how the apparent length of the baselines changes relative to the wavelength associated to the working frequency ($\lambda=c/f$). In addition, the top and bottom rows show the baselines from two different observation declinations, illustrating how the distribution of baselines in $uv$-space depends on the pointing direction.

    \subsection{Visibilities}
        For a given antenna configuration, operating frequency range, and source direction relative to the array, the $uv$-points define the effective aperture of the telescope in Fourier space. By sampling the observed Fourier sky ($I_{\rm sky}$) at the points $(u_i,v_i)$ we obtain the so-called visibilities (see \autoref{sec:interferometry}).
        If we represent the set of $uv$-points of the telescope aperture by a Fourier mask, we can describe the visibilities as in \autoref{eq:vis}, where $M$, the sampling mask, contains ones at $uv$-positions with samples and zeros elsewhere. Since \texttt{argosim} operates on two-dimensional images discretised on a regular pixel grid, the sampling mask must first be adapted to the same grid in order to sample the $uv$-points. The \texttt{imaging\_utils} module uses a nearest-neighbour gridding algorithm to assign $uv$-samples to the nearest pixel in the grid. By default, the $uv$-mask is binary, however it is possible to use the counts of $uv$-samples as weights for each visibility or to use a user-defined weighting scheme. Then, once gridded, the mask is multiplied with the Fourier sky model to obtain the visibilities. Finally, we add independent Gaussian noise to each visibility.

\begin{figure}
            \centering
            \includegraphics[width=\linewidth, trim={1.3cm 0 0cm 0},clip]{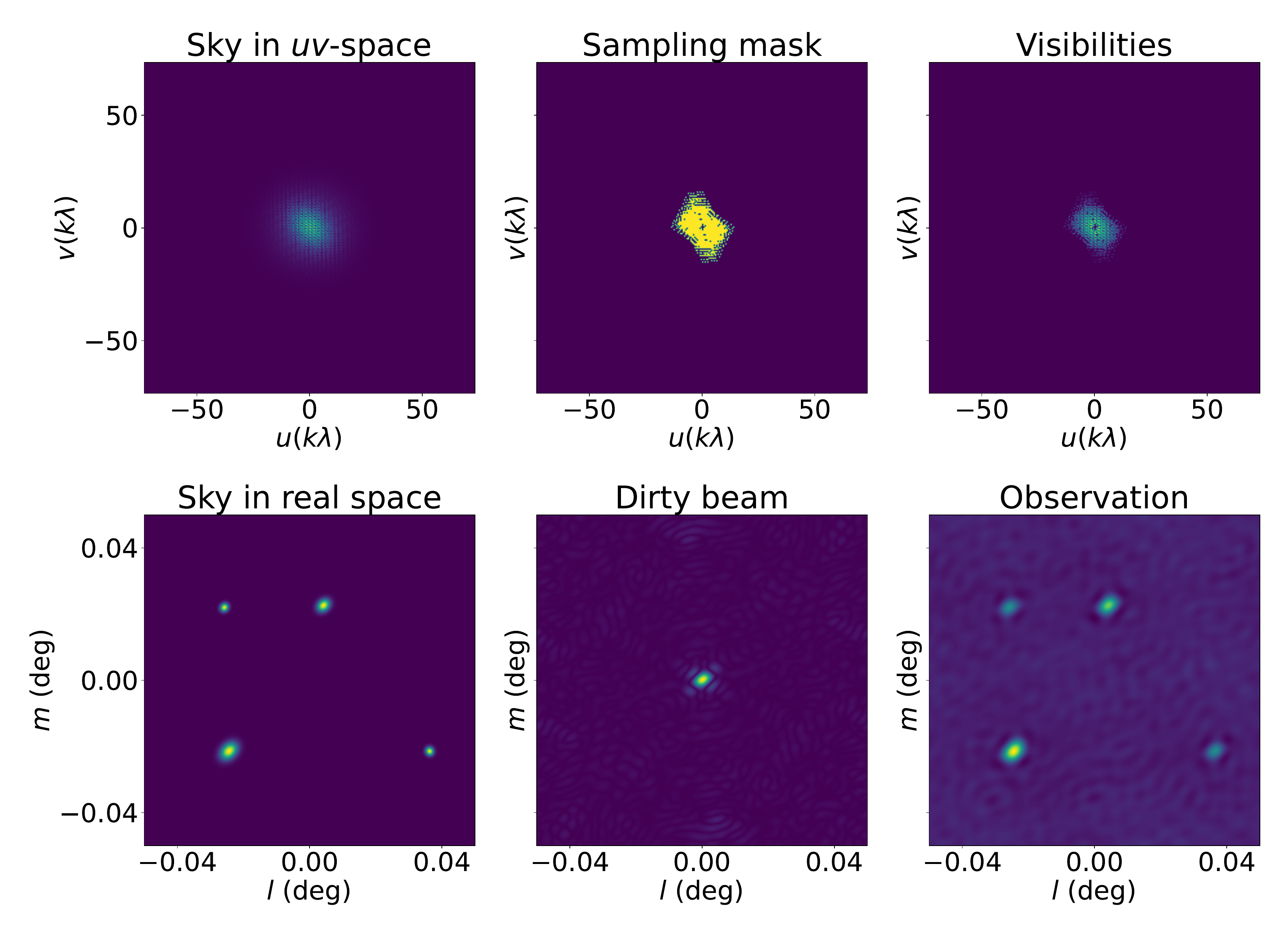}
            \caption{Radio interferometry observational model for four Gaussian sources. Top: sky model, $uv$-sampling mask and visibilities in Fourier space. Bottom: sky model, dirty beam and observation in real space. }
            \label{fig:obs_model}
        \end{figure}

        \Autoref{fig:obs_model} depicts the radio interferometry observational model. The top row shows the observation process in Fourier space, as described by \autoref{eq:vis}, showing the $uv$-space sky model, the gridded sampling mask and the gridded visibilities with additive noise. The bottom row shows the equivalent observation model in real space, defined in \autoref{eq:obs_model_real}, in which the real-space sky model, the dirty beam (PSF analogue) and the dirty observation with coloured noise are involved. The sky model used in this example is generated with \texttt{argosim} and consists of four Gaussian sources with non-diagonal covariance matrices, located at random positions in the FOV.

    \subsection{Sky model}
        The \texttt{data\_utils} module allows to generate simple sky models containing a number of sources. The sources are modelled by Gaussian profiles, which allow both point-like and resolved sources to be simulated. This morphology is relatively simple but is a good first-order approximation. Furthermore, software such as \texttt{PyBDSF} \citep{PyBDSF}, widely used in the radio astronomy community \citep{2025arXiv250611715V,2025AJ....170...60G,2025arXiv250620845L}, bases its blob detection and image decomposition on a set of Gaussian profiles. Each source is characterised by its shape, orientation, and size, defined by a two-dimensional covariance matrix, its flux, and its relative position in the FOV. 
        
        The \texttt{data\_utils} module provides an easy way to generate multi-source sky models of random shape, orientation, size and position. \Autoref{fig:obs_model} shows an example of a sky model generated with \texttt{argosim}, containing four sources of different flux, shape and size. In addition, \texttt{argosim} allows users to upload single channel images to be used as a custom sky model.
    
    \subsection{Image reconstruction}
        To recover the true sky intensity from the visibilities, different deconvolution algorithms can be used. The \texttt{clean} module currently applies the CLEAN algorithm \citep{1974A&AS...15..417H} to reconstruct the clean image. 
    
        CLEAN assumes that the sky is composed of point sources. It operates iteratively by identifying the position of the brightest source and subtracting a small fraction (defined by the gain) of its intensity, convolved with the dirty beam of the observation. This process is repeated until the peak intensity is smaller than a user-defined threshold or after a maximum number of iterations.
    
        Despite its simplicity, CLEAN enables the recovery of sources with reasonable accuracy but may introduce artefacts. Its main advantages are its speed, ease of implementation, and widespread adoption, making it a useful baseline for comparison. However, because is relies on the assumptions of point-like sources and sparse sky model, its performance is limited and future versions of \texttt{argosim} will include alternative reconstruction methods. Examples of CLEAN reconstructed images are shown in \autoref{sec:practical}, where we show a practical application of imaging in the context of the ARGOS pathfinder.
    
    \subsection{Metrics}
        
        The \texttt{metrics\_utils} module includes a set of imaging metrics for evaluating the quality of the reconstructed observation relative to the ground truth sky model, as well as a set of beam metrics for assessing the properties of the dirty beam and the $uv$-sample distribution.
        \subsubsection{Imaging metrics}
        The metrics provided to assess the quality of the reconstruction in real space are the mean squared error (MSE), the relative mean squared error (RMSE), the absolute pixel-wise residuals and the structural similarity index measure (SSIM). 
        The mean squared error metrics are computed as follows
        \begin{equation}
            {\rm MSE}(y, \hat{y}) := \frac{1}{N_{\rm px}} \sum_{\rm px}(y-\hat{y})^2,
        \end{equation}
        and
        \begin{equation}
            {\rm RMSE}(y, \hat{y}) := \frac{{\rm MSE}(y, \hat{y})}{{\rm MSE}(y, \bm{0})},
        \end{equation}
        where $\bm{0}$ represents the null matrix. The SSIM \citep{ssim} is provided by the image processing library \texttt{scikit-image} \citep{scikitimage}.
        
        \subsubsection{Beam metrics}
        The quality of a radio interferometric observation is largely determined by the shape and size of the dirty beam, which in turn depends on the distribution of samples in $uv$-space. To characterise the beam morphology, \texttt{argosim} provides a set of metrics for quantitative evaluation, defined in the \texttt{beam\_utils} module. One such tool is the \texttt{fit\_elliptical\_beam} function. As its name suggests, this function fits an ellipse to the main lobe of the dirty beam using the intensity thresholding method. This fit allows for the estimation of the beam size, in degrees or pixels, along the major and minor semi-axes. The ellipse is fitted at half the peak amplitude, corresponding to the full width at half maximum (FWHM) along each axis. The fit also provides the eccentricity and rotation angle of the main lobe, offering further insight into the morphology of the dirty beam.

        \begin{figure}
            \centering
            \includegraphics[width=\linewidth, trim={1.3cm 0 1.2cm 0},clip]{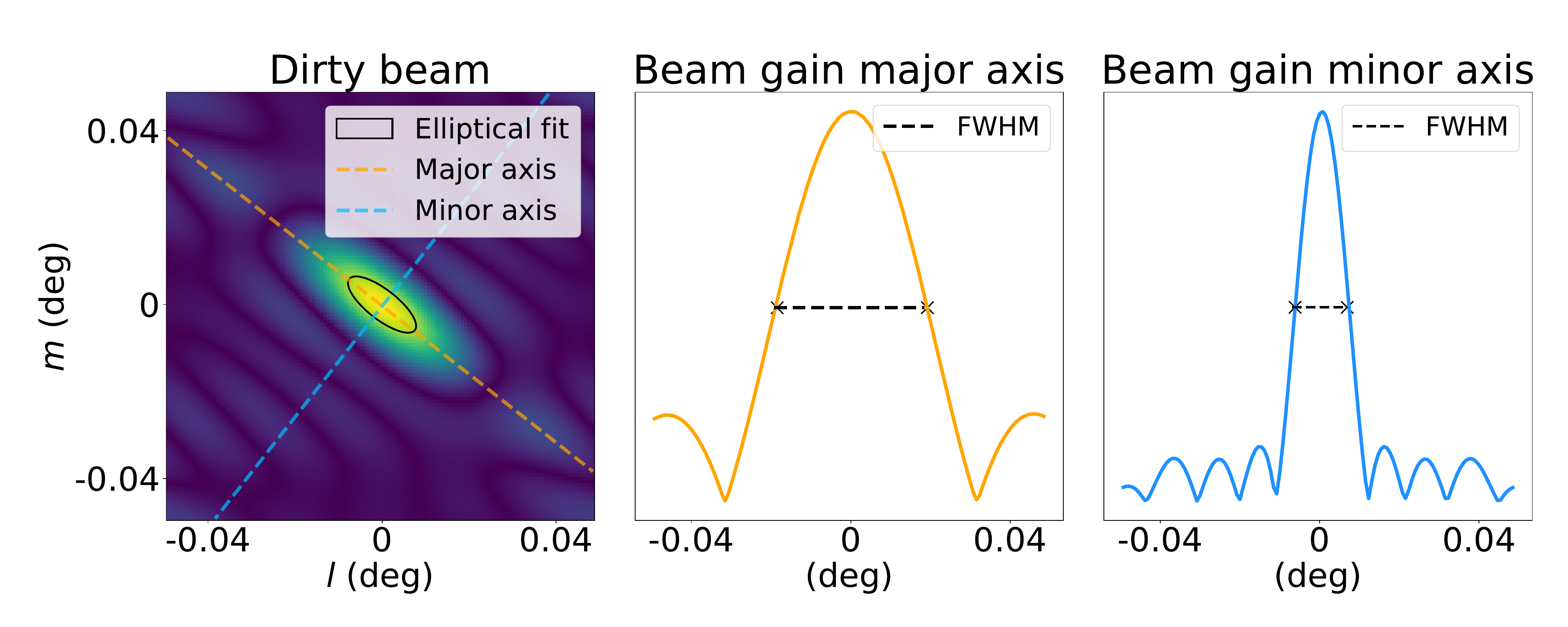}
            \caption{Left: dirty beam example with elliptical fit overlaid. The beam has an eccentricity of $0.93$ and is tilted at an angle of $-38$ degrees. Centre and right: the gain of the beam on the major and minor axes of the fit. Shown in black, the FWHM of the beam along the major axis ($0.039$ deg) and along the minor axis ($0.014$ deg).}
            \label{fig:beam_metrics}
        \end{figure}
        \Autoref{fig:beam_metrics} shows the dirty beam for a radio interferometer with 14 antennas distributed in a circle of 150 metres radius. Over the dirty beam, the elliptical fit is shown in black, as well as the major and minor axes of the ellipse in orange and blue, respectively. The eccentricity of the beam is $0.93$, where zero corresponds to a circle and the closer to one the more elongated the ellipse is, and the beam is tilted at an angle of $-38$ degrees.
        The two rightmost graphs show the dirty beam gain evaluated on the major and minor axes. The width of the beam at half maximum (FWHM) is also shown for each of the two directions. The FWHM along the major axis is $0.039$ degrees (equivalent to $49.6$ pixels) and along the minor axis is $0.014$ degrees (equivalent to $17.8$ pixels).
        
        Another highly relevant feature of the dirty beam is the side-lobe level (SSL), which is indicated on a logarithmic scale with respect to the maximum gain of the central lobe. The SSL is defined as follows
        \begin{equation}
            {\rm SSL}:= 10\; \log_{10}\left(\frac{b_{\rm sl}}{b_{\rm pk}}\right),
        \end{equation}
        where $b_{\rm sl}$ is the maximum gain of the side-lobe and $b_{\rm pk}$ is the peak of the main lobe. The SLL provides a quantitative measure of the amplitude of the largest side-lobes in the dirty beam. High side-lobes can lead to significant artefacts in the reconstructed image. A common challenge in radio interferometry arises when bright sources located just outside the field of view leak into the observation through these side-lobes, contaminating the image. Such contamination is particularly problematic, as it is difficult to identify and remove during post-processing.

    \subsection{Practical applications}
    \label{sec:practical}
    As demonstrated in this section, \texttt{argosim} enables the simulation of various antenna configurations, the computation of baselines and corresponding $uv$-sampling based on a given observation strategy, the synthesis and characterisation of the dirty beam using multiple metrics, and the simulation of interferometric observations. In the following, we present a practical example showcasing the application of \texttt{argosim} to simulate an observation with the ARGOS pathfinder telescope.

\begin{figure}
            \centering
            \includegraphics[width=\linewidth]{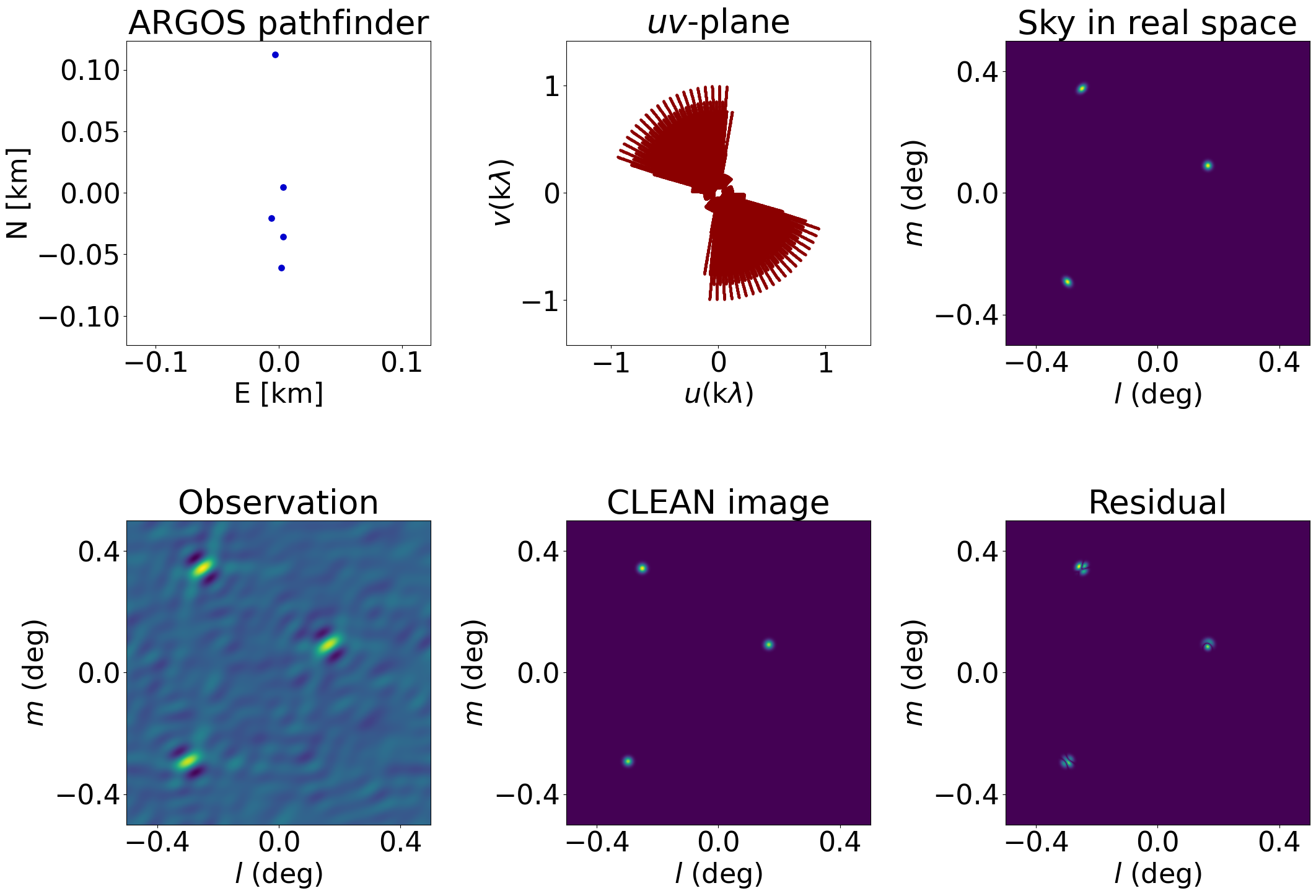}
            \caption{Practical application of imaging with \texttt{argosim}. The top row shows the array configuration and $uv$-sampling points for the ARGOS pathfinder array, as well as the observed sky model, composed of three Gaussian sources. The bottom row shows the observation, the CLEAN image and the residual between the reconstructed image and the sky model.}
            \label{fig:practical_app_argosCDS}
        \end{figure}

    The ARGOS pathfinder prototype consists of five six-metre diameter parabolic dishes distributed over an area of $20\times150$ metres. The array layout is shown in the top-left corner of \autoref{fig:practical_app_argosCDS}. The array latitude is $35^\circ$, which is the latitude of Heraklion (Crete, Greece), and the direction of observation is at a declination of $90^\circ$.
    We used \texttt{argosim} to calculate the sampling mask for a 5-hour observation with a timestep of 15 minutes, and 128 frequency channels equally spaced in the 1 to 3 GHz range. The $uv$-mask is shown in the top-middle panel of \autoref{fig:practical_app_argosCDS}.
    The observation is simulated for a FOV of $1 \deg^2$, using \texttt{argosim} to generate the sky model, which consists of three sources with randomly distributed Gaussian profiles. 
    The sky model has $512\times512$ pixels, that is, a pixel resolution of $7''$. The sky model is shown in the top-right panel of \autoref{fig:practical_app_argosCDS}.
    We then simulated the noisy observation of the sky model, which is shown in the lower left panel, and finally performed a basic deconvolution with the CLEAN algorithm. The clean image and the residual with respect to the ground truth sky model are presented in the lower middle and right panels.

    The quality of the reconstruction is quantitatively reported in \autoref{tab:imaging_metrics}, where we provide the image metrics (MSE, RMSE and SSIM) of the dirty observation and the clean observation with respect to the ground truth sky model. As expected in this simple observation case, the MSE and RMSE are significantly lower for the clean observation than for the dirty observation. The SSIM similarity metric also indicates a greater resemblance to the sky model in the case of the clean observation. Furthermore, in \autoref{tab:beam_metrics} we present the metrics of the synthesised beam of the ARGOS pathfinder for this observation case. The displayed metrics are the SSL, the FWHM over both semi-axes and the eccentricity of the main lobe. For this observation case, the ARGOS pathfinder beam has a SSL of $-3.4\;\rm dB$, which corresponds to an amplitude of $0.46$ relative to the maximum amplitude of the main lobe. The beam has a size of almost $20$ pixels along the major axis and $8$ pixels along the minor axis. This significant asymmetry is reflected in the eccentricity, which is much closer to one than it is to zero (perfect circle). The code used in this simulation example is available in our GitHub repository \footnote{\href{https://github.com/ARGOS-telescope/argosim/blob/main/notebooks/paper_figures_practical_application.ipynb}{notebooks/paper\_figures\_practical\_application.ipynb}}.

    \begin{table}
        \centering
        \caption{Imaging metrics for the ARGOS pathfinder simulation. Both dirty and clean observations are compared with the ground truth sky model.}
        \label{tab:imaging_metrics}
        \begin{tabular}{llll}
        Observation       & MSE    & RMSE  & SSIM \\ \hline\hline
        Dirty obs & $5.5 \times 10^{-4}$ & 81\%  & 0.64 \\
        Clean obs & $3.9 \times 10^{-5}$ & 5.7\% & 0.99
        \end{tabular}
    \end{table}

    \begin{table}
        \centering
        \caption{Metrics for the ARGOS pathfinder dirty beam. The displayed metrics are the SSL, the FWHM over both semi-axes and the eccentricity of the main lobe.}
        \label{tab:beam_metrics}
        \begin{tabular}{lll}
        SSL {[}dB{]} & FWHM {[}px{]} & Eccentricity \\ \hline\hline
        -3.4         & (19.6, 8.2)   & 0.91        
        \end{tabular}
    \end{table}

    \subsection{Graphical user interface (GUI)}
    To enhance user interaction with the \texttt{argosim} package, we provide a graphical user interface (\href{https://argos-telescope.github.io/argosim/running.html}{\texttt{argosim} GUI}\footnote{\url{https://argos-telescope.github.io/argosim/running.html}}) developed using \texttt{PyQt6} \citep{pyqt6}. The interface is structured into three main modules: one dedicated to antenna layout design, another focused on the aperture synthesis process, and a third addressing image reconstruction. The GUI offers full control over simulation parameters and enables real-time visualisation of their effects on $uv$-sampling, dirty beam formation, and imaging outcomes. Beyond improving usability, this interface serves as a valuable tool for outreach and education, offering an accessible platform for exploring key concepts in radio interferometry.


\section{Accelerated array computation}
\label{sec:jax}
    One of the key assets of \texttt{argosim} is that its imaging backbone is coded in \texttt{JAX} \citep{jax2018github}, a Python library for array-oriented numerical computation. \texttt{JAX} allows for significantly accelerated two-dimensional array operations such as matrix multiplication, matrix vector multiplication, and more complex operations such as the 2-Dimensional Fast Fourier Transform (2D-FFT). In addition of optimising these operations, highly present in the simulation of radio-interferometric observations, \texttt{JAX} allows through \texttt{vmap} (vectorising map) to generate vectorised implementations of functions automatically. Vectorisation enables functions to be executed on collections of inputs (1D array), thus avoiding sequential for loops.
    
\begin{figure}
        \centering
        \includegraphics[width=\linewidth]{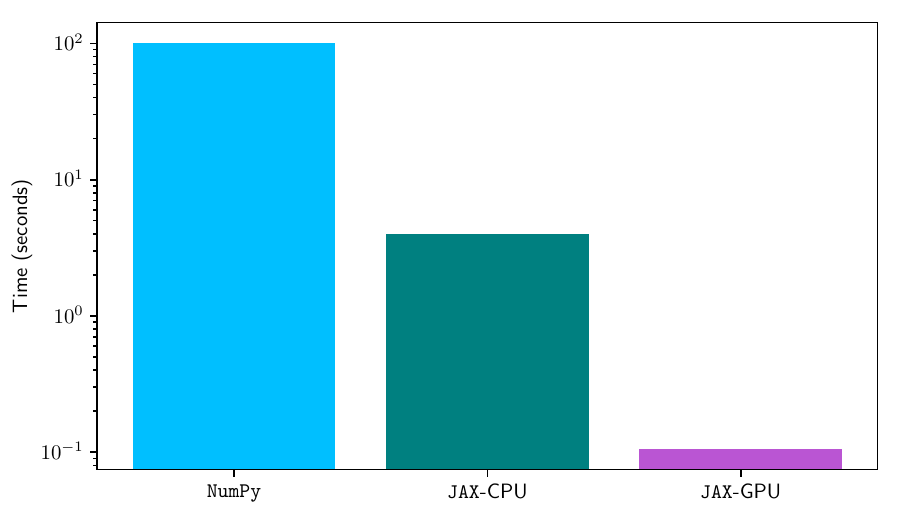}
        \caption{Simulation wall time for an interferometric observation using approximately 200 million $uv$-samples, compared across different computational backends.}
        \label{fig:performance_np_jax}
    \end{figure}

    \Autoref{fig:performance_np_jax} shows the total simulation time (wall time) for a radio interferometric observation using three computational backends: the \texttt{NumPy} implementation of \texttt{argosim} running on CPU (Apple M2 Pro 32 gb), the \texttt{JAX}-based version running on the same CPU, and the \texttt{JAX} implementation running on GPU (Nvidia Tesla V100 32gb). The simulation is carried out for a 27-element VLA-like array, which collects data for one hour with a timestep of 10 seconds, in 512 frequency bins between 1-3 GHz, adding up to a total of approximately 200 million $uv$-samples. We simulate a three-degree field of view, with a resolution of $2.64$ arcsec per pixel, that is, a $4\,096 \times 4\,096$ px$^2$ image. 

    As shown in \autoref{fig:performance_np_jax}, the \texttt{JAX} reimplementation, executed on the same CPU as the \texttt{NumPy} version, achieves a simulation speedup exceeding $10 \times$. Remarkably, the \texttt{JAX} implementation on GPU delivers a substantial acceleration of nearly $1\,000 \times$ compared to the original \texttt{NumPy} case. This computational acceleration is essential in the era of big data, particularly for processing the vast volumes of data expected from next-generation interferometers.

    In addition to improved performance, \texttt{JAX} enables the development of a fully differentiable simulation pipeline. By ensuring that each step in the observation process is differentiable, \texttt{JAX} can track computations and automatically propagate gradients via automatic differentiation (\textit{autodiff}). Such a differentiable pipeline is particularly useful in gradient-based optimisation tasks and in posterior sampling algorithms, such as Hamiltonian Monte Carlo (HMC).
    
\section{Future releases}
\label{sec:future}

Future development of \texttt{argosim} will include both minor releases for bug fixes and incremental improvements, as well as major releases introducing substantial new features. Planned enhancements include the integration of realistic sky simulations using \texttt{GalSim} \citep{rowe2015galsimmodulargalaxyimage}, support for multi-band imaging through spectral sky model cubes (moving beyond the current flat-spectrum assumption), and the development of more advanced deconvolution algorithms. Additional functionality under consideration includes more sophisticated gridding and de-gridding methods to improve visibility sampling accuracy, and advanced noise modelling capabilities such as radio frequency interference (RFI) simulation.
Finally, we are further developing the \texttt{argosim} GUI to enhance user experience and to leverage its potential as a pedagogical tool and a platform for rapid experimentation.

\section{Conclusion}
\label{sec:conclusions}
In this paper, we presented the first public release of the open-source Python package \texttt{argosim}, a toolkit designed for radio interferometric simulations. The package is lightweight, modular, easy to install, and leverages \texttt{JAX} for accelerated computation. We described the core modules of \texttt{argosim} and outlined the complete pipeline for simulating radio interferometric observations.

The \texttt{argosim} package enables the modelling of various array configurations, including user-defined layouts. It supports the computation of baselines and the generation of the corresponding $uv$-sampling mask based on the selected observation strategy. Simulations can include multiple frequency bands and account for time integration effects due to the rotation of the Earth. The resulting sampling mask is used to compute the visibilities of a given sky model, with optional noise inclusion in $uv$-space. 
For sky modelling, \texttt{argosim} includes functionality for generating synthetic skies composed of multiple Gaussian sources and also supports user-provided sky models. Image reconstruction is performed using an implementation of the CLEAN deconvolution algorithm, and the quality of the reconstruction can be quantitatively assessed using the metrics module. This module also provides a detailed characterisation of the dirty beam, including its shape, size, and side-lobe level.
Finally, we demonstrated a practical application of \texttt{argosim} by simulating an observation with the ARGOS pathfinder telescope. We computed the $uv$-sampling mask for a 5-hour observation over a 2 GHz bandwidth, simulated the visibilities, applied CLEAN deconvolution, and evaluated both the image reconstruction and beam metrics.

We invite users and the wider community to provide feedback on this first release of \texttt{argosim}. Issues related to installation, bugs, or feature requests can be submitted directly through the project's GitHub repository. As an open-source initiative, \texttt{argosim} welcomes contributions from all interested developers and researchers. New features and improvements can be proposed via pull requests, fostering a collaborative and transparent development process.

\begin{acknowledgements}
      This work was supported by the European Community through the grant ARGOS (contract no. 101094354). 
      This work was granted access to the HPC resources of IDRIS under the allocation 2025-AD010414104R2 made by GENCI. We acknowledge the members of the ARGOS consortium for their valuable feedback on \texttt{argosim}.
\end{acknowledgements}

\bibliographystyle{aa}
\bibliography{astro}

%

\begin{appendix}
\onecolumn

\section{Third-party software}  \label{app:dependencies}
The third-party software used in \texttt{argosim} are listed in \autoref{tab:dependencies} with the associated versions, and references. The table includes the packages required for running the GUI, building the documentation, formatting the code and testing \texttt{argosim}.
\begin{table}[ht]

  \caption{Third-party software used in \texttt{argosim}.}
\label{tab:dependencies}
\begin{tabular}{lll}
\hline
\hline
    Package name & Version & References
    \\ \hline
    \multicolumn{3}{c}{\textbf{\texttt{argosim}}} \\
    \texttt{JAX} & $0.4.25$ & \cite{jax2018github}\\ 
    \texttt{Matplotlib} & $\geq 3.8$ & \cite{Hunter_2007} \\ 
    \texttt{NumPy} & $\geq 1.26$ & \cite{Harris_2020}  \\
    \texttt{scikit-image} & $\geq 0.22$ & \cite{scikitimage} \\
    \hline \\
    \multicolumn{3}{c}{\textbf{\texttt{argosim}[gui]}} \\
    \texttt{PyQt6} & $\geq 6.9.1$ & \cite{pyqt6}\\
    \hline \\
    \multicolumn{3}{c}{\textbf{\texttt{argosim}[doc]}} \\
    \texttt{Sphinx} & $\geq 8.1.3$ & \cite{sphinx} \\
    \hline \\
    \multicolumn{3}{c}{\textbf{\texttt{argosim}[lint]}} \\
    \texttt{Black} & $\geq 25.1.0$ & \cite{black} \\
    \texttt{isort} & $\geq 5.13.2$ & \cite{isort} \\
    \hline \\
    \multicolumn{3}{c}{\textbf{\texttt{argosim}[test]}} \\
    \texttt{pytest} & $\geq 8.3.3$ & \cite{pytest} \\
    \hline
  \end{tabular}

\end{table}

\clearpage

\section{\texttt{argosim} modules} \label{app:modules}
The main modules and functions available in \texttt{argosim} that serve the pipeline described in \autoref{sec:pipeline} are listed in \autoref{tab:modules}. More details can be found in our extensive documentation\footnote{\url{https://argos-telescope.github.io/argosim/}}.
\begin{table}[ht]
    \renewcommand*{\arraystretch}{1.2}
  \caption{List and brief description of the \texttt{argosim} modules, functions and classes.}
  \label{tab:modules}
    \begin{tabularx}{\textwidth}{lX}
    \hline
    \hline
    Name & Description
    \\ \hline
    \multicolumn{2}{c}{\textbf{\texttt{antenna\_utils} module}} \\ [1mm]
    \texttt{ENU\_to\_XYZ} & Convert the baselines from ENU to XYZ coordinates. \\ 
    \texttt{XYZ\_to\_uvw} & Get the $uvw$ sampling points from the XYZ coordinates given the observing parameters.  \\ 
    \texttt{circular\_antenna\_arr} & Generate a circular antenna array. \\ 
    \texttt{combine\_antenna\_arr} & Combine two antenna arrays. \\ 
    \texttt{get\_baselines} & Compute the baselines of an antenna array. \\  
    \texttt{load\_antenna\_enu\_txt} & Load antenna name, ENU positions and noise from a text file. \\ 
    \texttt{load\_antenna\_latlon\_txt} & Load the antenna name, latitude, longitude, altitude and noise from a text file.  \\ 
    \texttt{random\_antenna\_arr} & Generate a random antenna array.  \\ 
    \texttt{random\_antenna\_pos} & Generate a random antenna location in ENU coordinates. \\ 
    \texttt{save\_antenna\_enu\_txt} & Save the antenna name, ENU positions and noise into a text file. \\ 
    \texttt{uni\_antenna\_array} & Generate a uniform antenna array.  \\ 
    \texttt{uv\_track\_multiband} & Compute the $uv$-sampling baselines for a given observation time and frequency range. \\ 
    \texttt{y\_antenna\_arr} & Generate a Y-shaped antenna array. \\
    \hline
    \multicolumn{2}{c}{\textbf{\texttt{beam\_utils} module}} \\ [2mm]
    \texttt{CosCubeBeam} & Class to model the primary beam of the antennas using a cosine cubed function. \\ 
    \hline
    \multicolumn{2}{c}{\textbf{\texttt{clean} module}} \\ [1mm]
    \texttt{clean\_hogbom} & Perform the Högbom’s CLEAN algorithm.  \\ 
    \texttt{find\_peak} & Find the peak of an image. \\ 
    \texttt{shift\_beam} & Shift the beam image by a given amount of pixels. \\
    \hline
    \multicolumn{2}{c}{\textbf{\texttt{data\_utils} module}} \\ [1mm]
    \texttt{gauss\_source} & Generate a 2D Gaussian source. \\
    \texttt{n\_source\_sky} & Generate a sky image with multiple Gaussian sources at random positions. \\ 
    \texttt{random\_source} & Generate 2D Gaussian source with random mean and covariance. \\ 
    \hline
    \multicolumn{2}{c}{\textbf{\texttt{imaging\_utils} module}} \\ [1mm]
    \texttt{add\_noise\_uv} &
    Add white Gaussian noise to the visibilities in the $uv$-plane. \\
    \texttt{compute\_visibilities\_grid} & Compute the visibilities from the Fourier sky and the $uv$-sampling mask. \\
    \texttt{grid\_uv\_samples} & Compute the discrete $uv$-sampling mask from the continuous $uv$-sampling points.  \\ 
    \texttt{simulate\_dirty\_observation} &  Simulate a radio observation of the sky model from the track $uv$-samples. \\ 
    \texttt{sky2uv} & Compute the Fourier transform of the sky. \\ 
    \texttt{uv2sky} & Compute the inverse Fourier transform of the $uv$-plane. \\
    \hline
    \multicolumn{2}{c}{\textbf{\texttt{metrics\_utils} module}} \\ [1mm]
    \texttt{compute\_eccentricity} & Compute the eccentricity of the beam main lobe. \\ 
    \texttt{compute\_fwhm} & Compute the FWHM from the beam. \\
    \texttt{compute\_sll} & Compute the SLL of a beam using an elliptical mask. \\ 
    \texttt{fit\_elliptical\_beam} & Fit an ellipse to the main lobe of the dirty beam. \\
    \texttt{mask\_main\_lobe\_elliptical} & Apply an elliptical mask to suppress the main lobe from a beam image. \\ 
    \texttt{mse} & Compute the mean squared error between two images. \\ 
    \texttt{residuals} & Compute the residuals between two images. \\
    \texttt{ssim} & Compute the structural similarity index between two images. \\
    
    \hline
  \end{tabularx}
\end{table}

\FloatBarrier 
\clearpage

\end{appendix}
\end{document}